\documentclass[notitlepage,twocolumn,superscriptaddress,amsmath,amssymb,aps,prl]{revtex4-1}
\pdfoutput=1
\usepackage[utf8]{inputenc}
\usepackage[T1]{fontenc}
\usepackage{hyperref}
\usepackage{graphicx}
\usepackage{amsfonts, amsmath, amsthm, amssymb} 
\usepackage{braket}
\usepackage{float}
\usepackage{amsthm}
\usepackage{color}
\usepackage[all]{nowidow}
\usepackage[dvipsnames]{xcolor} 

\usepackage{csquotes}

\hypersetup{citecolor=black,colorlinks=false,urlcolor=black} 
\usepackage[english]{babel}

\newcommand{\tr}[1]{\text{tr}\left(#1\right)}
\newcommand{\lindblad}[1]{\mathcal{L}\left(#1\right)}
\newcommand{\identity}{\mathbb{I}}

\newcommand{\change}[1]{\textcolor{black}{#1}}
\newcommand{\changenew}[1]{\textcolor{black}{#1}}

\begin{document}

\title{Certifying steady-state properties of open quantum systems}

\author{Luke Mortimer}
\affiliation{ICFO-Institut  de  Ciencies  Fotoniques,  The  Barcelona  Institute  of  Science  and  Technology, 08860 Castelldefels (Barcelona), Spain}

\author{Donato Farina}
\affiliation{Physics Department E. Pancini - Università degli Studi di Napoli Federico II, Complesso Universitario Monte S. Angelo - Via Cintia - I-80126 Napoli, Italy}

\author{Grazia Di Bello}
\affiliation{Physics Department E. Pancini - Università degli Studi di Napoli Federico II, Complesso Universitario Monte S. Angelo - Via Cintia - I-80126 Napoli, Italy}

\author{\\ David Jansen}
\affiliation{ICFO-Institut  de  Ciencies  Fotoniques,  The  Barcelona  Institute  of  Science  and  Technology, 08860 Castelldefels (Barcelona), Spain}

\author{Andreas Leitherer}
\affiliation{ICFO-Institut  de  Ciencies  Fotoniques,  The  Barcelona  Institute  of  Science  and  Technology, 08860 Castelldefels (Barcelona), Spain}

\author{Pere Mujal}
\affiliation{ICFO-Institut  de  Ciencies  Fotoniques,  The  Barcelona  Institute  of  Science  and  Technology, 08860 Castelldefels (Barcelona), Spain}

\author{Antonio Ac\'{i}n}
\affiliation{ICFO-Institut  de  Ciencies  Fotoniques,  The  Barcelona  Institute  of  Science  and  Technology, 08860 Castelldefels (Barcelona), Spain}
\affiliation{ICREA-Institucio Catalana de Recerca i Estudis Avan\c cats, Lluis Companys 23, 08010 Barcelona, Spain}

\date{\today}

\begin{abstract}
\noindent

\change{Estimating the steady-state properties of open many-body quantum systems is a fundamental challenge in quantum science and technologies. In this work, we present a scalable approach based on semi-definite programming to derive certified bounds on the expectation value of an arbitrary observable in the steady state of Lindbladian dynamics.} We illustrate our method on a series of many-body systems, including \change{paradigmatic spin-1/2} chains and two-dimensional ladders, \change{considering both equilibrium and nonequilibrium steady-states.} We benchmark our method with state-of-the-art tensor-network approaches \change{that, unlike our method, are only able to provide estimates, with no guarantee, on steady-state quantities}. For the tested models, only modest computational effort is needed to obtain certified non-trivial bounds for system sizes intractable by exact methods.
\change{Our method introduces the first general numerical tool for bounding steady-state properties of open quantum dynamics, opening a new avenue in the understanding of stable configurations in many-body systems.}

\end{abstract}

\maketitle

{\it Introduction.---}
Open quantum systems, which interact continuously with their surrounding environment, are fundamental to understanding a vast array of physical phenomena and technological applications, including quantum optics, condensed matter physics, and quantum information processing \cite{carmichael2009open,rivas2012open,breuer2002theory}. These interactions often lead to dissipative dynamics that can significantly alter the system's behavior over time. 
Accurately determining the steady-state properties of such systems is crucial for predicting long-term behavior and for the design of quantum devices that can operate reliably under real-world conditions \cite{lieu2020symmetry,landi2020thermodynamic}.
From a general perspective, there has been increasing recent interest \cite{fazio2024many,ilin2023learning} in finding techniques to compute 
the stationary state and its properties \cite{rota_critical_2017,PhysRevLett.130.240601,morrone_estimating_2024}, especially with variational methods~\cite{nagy_driven-dissipative_2018, vicentini_variational_2019, nagy_variational_2019, hryniuk2024tensor}, in characterizing multiple steady states~\cite{thingna_degenerated_2021,amato_number_2024}
and in designing Lindblad generators for target steady states \cite{guo_designing_2024, souza_lindbladian_2024}.
On the experimental side, a growing effort was devoted to demonstrate 
stable quantum-correlated many-body states \cite{mi_stable_2024},
steady-state superradiance in free space \cite{ferioli_non-equilibrium_2023}
and propose Dicke superradiant enhancement of heat current in circuit quantum electrodynamics \cite{andolina_dicke_2024}.

A common framework for modeling the dynamics of open quantum systems is through the Lindblad master equation \cite{gorini1976completely,lindblad1976generators},
\begin{equation}
\dot \rho=\lindblad{\rho}\,,
\end{equation}
which provides a Markovian description of the time evolution of the system's density matrix 
\cite{breuer2002theory, manzano2020short}. The Lindbladian superoperator \changenew{has the form (in natural units} \cite{hbar}),
\begin{equation}
    \lindblad{\rho} = -{i}[H,\rho ]+\sum _{i}^{}\gamma _{i}\left(L_{i}\rho L_{i}^{\dagger }-{\frac {1}{2}}\left\{L_{i}^{\dagger }L_{i},\rho \right\}\right)\,,
    \label{eq:Lindblad}
\end{equation}
which encapsulates both the unitary evolution governed by the system's Hamiltonian \change{$H$} and the dissipative processes arising from environmental interactions. Finding the steady state, $\rho_{ss}$, involves solving for the density matrix that remains static under the Lindbladian dynamics,
\begin{equation}
    \lindblad{\rho_{ss}} = 0\,.
    \label{ss-eq}
\end{equation}
The exact computation of the steady state is possible \change{for small systems (in the case of qubits, number of qubits $n\lesssim 8$ \cite{end-matter-app-dm})},
but soon becomes intractable when increasing the system size due to exponential scaling. In fact, the situation is harder than what is observed for other relevant many-body problems, such as the computation of ground states, as here one has to deal with mixed states. To reach larger sizes, one usually employs approximate variational approaches, such as tensor network \change{(TN)} methods~\cite{jaschke2018one}. 
Unfortunately, \changenew{these methods only provide an estimate for the steady state, without guarantees on how close its observables are to the exact values. Instead, the method in this work allows deriving upper and lower bounds that determine an interval within which the true value lies with absolute certainty.}

In many situations, one does not need a full description of the steady state as one is only interested in knowing a few relevant observables. Determining the complete state may in fact be an unnecessarily complex task if, in the end, one is only going to be able to compute a much smaller, say polynomially growing in the system size, number of parameters. It may then be more practical to search for methods that target the observable(s) of interest without requiring the full reconstruction, or approximation, of the steady state. In this work, we follow this approach and provide a scalable method to derive bounds on expectation values of any observable of interest at the steady state of open-system dynamics described by a Lindblad master equation. To do so, we exploit relaxations for polynomial optimization introduced in the context of quantum information theory~\cite{navascues2007bounding,navascues2008convergent,PNA10} which make use of semi-definite programming (SDP). We apply our approach to paradigmatic models and compare them with \change{TN} methods, seeing that they attain similar performance but with certificates, something difficult, if not impossible, to obtain in variational methods~\cite{wu2023variational}. Furthermore, while variational methods generally require the steady state to be unique \cite{hryniuk2024tensor}, our approach also applies beyond this instance providing bounds over the whole convex set of steady states.

We are thus interested in bounding the expectation value of an observable $O$ for the steady state of the system, i.e. minimizing/maximizing $\tr{O \rho}$ when $\lindblad{\rho} = 0$. Since this is linear in $\rho$, combined with the fact that a density matrix should be positive and have trace one, one can quite naturally formulate such a problem as an SDP \change{\cite{end-matter-app-dm}}. The full optimization, however, requires an explicit form of $\rho$, which scales exponentially with the number of constituents, $n$, typically qubits.

To avoid the exponential growth, in what follows we restrict positivity constraints to moment matrices made of subsets of all possible $n$-qubit operators. While the operators to construct moment matrices are arbitrary, a very natural choice for $n$-qubit systems consists of the ``Pauli strings'', made of tensor products of Pauli matrices and the identity operator for each qubit.
In this work, we will use the notation as follows: Pauli strings are denoted by $P_{\vec\alpha}$, where $\vec\alpha$ is a vector of $n$ components that can take values in $\{0,1,2,3\}$. For instance, $P_{310}$ refers to $\sigma_z \otimes \sigma_x \otimes \identity$, or even simpler $Z_1 X_2$. \change{Such Pauli strings have an ``order'' equal to the number of non-identity operators in the string (for example, the previous Pauli string has order 2).} Any generic observable can be written as a linear combination of Pauli strings with known real coefficients, for instance, in the case of our objective, $O=\sum_{\vec\alpha} o_{\vec\alpha} P_{\vec\alpha}$.

In what follows, we provide several constraints that can be used to bound expectation values of steady-state observables through convex \change{optimization techniques \cite{boyd2004convex}}.

{\it {Moment matrix positivity constraints}.---}
%
For a given state $\rho$ and vector of \change{$s\geq 1$} operators \change{$(\theta_i )_{i=1}^s$}, we can define $M$ as the corresponding moment matrix with elements 
\begin{equation}
    M_{ij} := {\rm tr}(\rho \theta_i^\dag \theta_j). \label{eq:def_momentmatrix}
\end{equation}
It can be shown~\cite{navascues2007bounding} that such a moment matrix is positive semidefinite \change{for any $s$ and choice of operators $(\theta_i)_{i=1}^s$.}
As an example of such a moment matrix constraint we can consider $M$ as in
{\eqref{eq:def_momentmatrix} with \change{ $s=7$, $(\theta_i)_{i=1}^s \equiv (\identity, {X_1}, {Y_1}, {Z_1}, {X_2}, {Y_2}, {Z_2})^T$}, and impose $M\succcurlyeq 0$.} We refer to this as the level 1 moment matrix, as the top row contains only first-order expectation values. The level 2 moment matrix would contain all first- and second-order expectation values as the top row, and so on. 
Here we also use the commutation rules between different \changenew{spins} (\change{e.g.}, $[X_i, X_j] = 0$ for $i\neq j$) as well as the Pauli reduction rules ($X_i X_i = \identity$, $X_i Y_i = i Z_i$, etc.) to simplify the matrix. In general, the size of moment matrix of level $k$ grows like $n^k$, polynomially with the number of qubits.

{\it {State reconstruction positivity constraints}.---}
Whilst it is completely \change{intractable} to constrain the positivity of the whole density matrix, what remains feasible is imposing positivity of all the reduced density matrices of a given, smaller, number of qubits, $m\ll n$. These positivity constraints are a function of the Pauli strings acting on the corresponding $m$ \changenew{spins}. By reconstructing all $m$-site density matrices and enforcing their positivity, we constrain the set of feasible moments and improve the tightness of our bounds, often significantly, at minimal computational cost. 

{\it Linear Lindbladian constraints.---}
Whilst moment matrix optimization has been well documented in the literature \cite{kogias2015hierarchy,pozas2019bounding}, including the case of bounding observables for many-body ground states \cite{wang2024certifying}, the case of steady-state optimization offers an extra set of constraints. Based on the definition of the steady state, we have the extra constraint from Eq.\,\eqref{ss-eq} that the Lindbladian acting on our state should be zero.
We can then use the adjoint Lindbladian \cite{breuer2002theory} in order to derive constraints in terms of moments of observables acting on $\rho$. Considering a generic operator $G$ and the definition of adjoint Lindbladian $\mathcal{L}^\dag$ we have that
\begin{eqnarray}
&&0=\tr{G \mathcal{L}(\rho_{ss}) } = \tr{\mathcal{L}^\dagger(G) \rho_{ss}} =\braket{\mathcal{L}^\dagger(G)}_{ss}\,,
\label{linear-contr}
\\&& \mathcal{L}^\dag(G) = {i}[H,G ]+\sum _{i}^{}\gamma _{i}\left(L_{i}^{\dagger }G L_{i}-{\frac {1}{2}}\left\{L_{i}^{\dagger }L_{i},G \right\}\right)\,.
    \nonumber
\end{eqnarray}
The result is an equation \change{into} which we can place any operator to give us a linear constraint on the system.
There is at least one steady state for a master equation in Lindblad form \cite{breuer2002theory}.
\change{\change{ In the case of a single steady state, }placing all possible $4^n-1$ nontrivial Pauli strings (i.e., all except the identity) provides a fully constrained system of $4^n$ independent linear equations when combined with the trace-one condition on the state, $\langle I \rangle_{ss}=1$, which can then be solved to retrieve the steady state from its Pauli decomposition.}
As anticipated, such a problem is only solvable at small values of $n$, say $n\lesssim 8$. In what follows, we will consider scalable relaxations involving only a subset of these linear constraints.

{\it \enquote{Automatic} constraint generation.---}
The previous ideas relax the exact treatment by restricting the positivity or Lindbladian constraints of the unknown steady state to a subset of Pauli strings. The choice of these strings is arbitrary, but\change{, with the \change{goal} of achieving tighter bounds with as few variables as possible, we choose to apply the following algorithm to generate a subset of Pauli strings that should be relevant to our system and observable. We begin by placing the Pauli strings from the decomposition of the observable of interest $O$ into the adjoint Lindbladian, as in Eq.\,\eqref{linear-contr}, to generate new linear constraints.}
We then take the \change{Pauli strings} that appear in the resulting \change{constraints} and plug them back into the adjoint Lindbladian. This process is repeated \change{until we have generated a sufficiently large set of Pauli strings and corresponding linear constraints}. In some cases we note that this eventually reaches a ``cycle'', in that no new operators are generated by the insertion in Eq.\,\eqref{linear-contr} of the previous operators. In this case, the resulting constraints \changenew{can} form a fully constrained linear system, which can then be solved as before. The set of Pauli strings resulting from the whole iterative process is in turn also used to construct the moment matrix. \change{
We refer to this method of selecting a subset of informative Pauli strings as \enquote{automatic}. This approach \change{appears} more effective than, e.g., using randomly chosen Pauli strings or considering all Pauli strings of a given order.}


{\it {Bounding expectation values of steady-state observables}.---}
By following the previous recipe, a subset of Pauli strings $S_M$ is identified, which contains all those needed to specify the observable and the different constraints. We can now formulate our optimization problem in its general form: \change{the lower/upper bound is given by}
\begin{equation}
\label{sdp-notation}
\begin{aligned}
\change{\langle O\rangle_{\rm lb/ub}=}
\underset{\change{\{\braket{P_{\vec\alpha}} \}_{\vec \alpha \in S_M}}}{\change{\text{min/max}}}\, &\sum_{\vec\alpha} o_{\vec\alpha} \braket{P_{\vec\alpha}} \\
\text{s.t.} \,&-1\leq \braket{P_{\vec\alpha}} \leq 1\,,\, \forall \vec\alpha\in S_M \\
& \braket{ \mathcal{L}^{\dagger}(P_{\vec\alpha})} =0\,,\, \forall \vec\alpha\in S_L  \\
& \sum_{\vec\alpha} A_{k,\vec\alpha} \braket{P_{\vec\alpha}} \succcurlyeq 0 \,,\, \forall k \in \{1, \dots,m_p\} \\
& \sum_{\vec\alpha} b_{k,\vec\alpha} \braket{P_{\vec\alpha}} = 0 \,,\, \forall k \in \{1,\dots,m_s\} \,.\\
\end{aligned}
\end{equation}
%
\change{ These two SDPs define an interval \begin{equation}
\label{lbub}
    \langle O \rangle_{\rm lb}\leq \langle O\rangle_{ss}\leq \langle O \rangle_{\rm ub}
\end{equation}
excluding the possibility that $\langle O\rangle_{ss}$ lies outside it.}
The bounds \enquote\change{$-1\leq \braket{P_{\vec\alpha}} \leq 1,\, \forall \alpha \in S_M$} are not always necessary, but provide a ``safety net'' to prevent the problem being unbounded in the case that one of the moments is not sufficiently constrained. The set $S_L$ identifies those Pauli strings used in the linear constraints. The $m_p$ linear matrix inequality constraints defined by known matrices $A_{k,\alpha}$ can correspond either to moment matrices
(obeying some Pauli replacement rules) or to reconstructed reduced density matrix positivity \cite{end-matter-app-dm}. Finally, the last $m_s$ linear equality constraints defined by coefficients $b_{k,\alpha}$ represent \change{known} symmetries specific to the problem. Such constraints add negligible computational cost, but often improve the bounds by a non-negligible amount \cite{suppmat}. Our code implementing the method is publicly available \cite{code}.\nocite{alicki2018introduction,levy2014local}

\begin{figure}[t]
    \centering
    \includegraphics[width=0.35\textwidth]{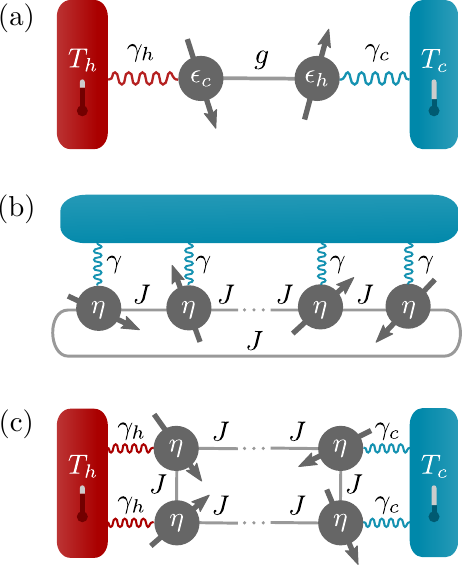}
    \caption{(a) 
    Two-qubit example: Here the left qubit is connected to a hot bath and the right qubit is connected to a cold bath. The system is defined by parameters $\gamma_h$, $\gamma_c$, coupling of the qubits with their respective baths, $\epsilon_h$, $\epsilon_c$, energy gaps of the qubits, $T_c$, $T_h$, the temperatures of the baths, and $g$, coupling between the two qubits.
    (b) 
    1D periodic chain example: Here each qubit is coupled to the dissipative bath with coefficient $\gamma$ and coupled to its nearest neighbours with coefficient $J$. The final qubit is also coupled to the first qubit. A field of coefficient \change{$\eta$} is also applied to each qubit in the $X$ direction.
    (c) 
    2D ladder example: Here the end qubits are connected to the respective heat bath with coefficient $\gamma_h$ \change{ and} $\gamma_c$, and \change{ each qubit is} coupled to its nearest neighbours with coefficient $J$. A field of coefficient \change{$\eta$} is also applied to each qubit in the $Z$ direction.}
    \label{fig:merged}
\end{figure}

{\it {Two-qubit test case}.---}
We first consider the simple test case discussed, e.g., by Refs.\,\cite{hofer2017markovian, khandelwal2020critical, PhysRevE.109.034112}:
%
\change{a two-qubit system, in which the left qubit is connected to a hot bath and the right qubit is connected to a cold bath, as in Fig.\,\ref{fig:merged}(a).} 
Choosing the local master equation, 
the open-system dynamics is described by the following Lindbladian,
\begin{equation}
  \lindblad{\rho}=-i\left[H, \rho\right] +\sum_{j \in\{h, c\}} ( \gamma_j^{+} {D}[\sigma_{+}^{(j)}] \rho+\gamma_j^{-} D[\sigma_{-}^{(j)}] \rho ) \,, 
  \label{eqn:2qubitlimblad}
\end{equation}
where
$
D[A] ~ B := A B A^\dagger -
\frac{1}{2}\{A^\dagger A , B\}\,.
$ 
\change{The Hamiltonian is $H:=H_{S} +H_{\text{int}}$, where
 $H_{S}=\sum_{j \in\{h, c\}} \varepsilon_j \sigma_{+}^{(j)} \sigma_{-}^{(j)}$ and  $H_{\text{int}}=g(\sigma_{+}^{(h)} \sigma_{-}^{(c)}+ \sigma_{-}^{(h)} \sigma_{+}^{(c)})$\,,
with
$\epsilon_h$ and $\epsilon_c$ being the energy gaps of the qubits, and $g$ the coupling constant. Here $\sigma_{+}^{(j)}=\frac{1}{2}(X_j+iY_j)$ and $\sigma_{-}^{(j)}=\frac{1}{2}(X_j-iY_j)$.}
The kinetic coefficients 
$ 
\gamma_j^{+}=\gamma_j n_{B}^j$ and $ \gamma_j^{-}=\gamma_j(1+n_{B}^j)\,
$ 
account, respectively, for absorption from and emission to bath $j$ and 
$ n^j_B = 1 / (e^{\epsilon_j / T_j} - 1) 
$
are the Bose factors.
Furthermore,
$\gamma_h$ and $\gamma_c$ are the couplings of the qubits with their respective baths \change{ at temperatures $T_h$ and $T_c$}.
In \cite{khandelwal2020critical} the authors focus on the 
heat current \change{ in the nonequilibrium steady state (NESS)} defined as $J_{ss} := Q_h^{ss}-Q_c^{ss}$, with
\begin{equation}
\label{heat-khandel}
  Q_j^{ss} = \operatorname{tr}\left[H_S\left(\gamma_j^{+} D\left[\sigma_{+}^{(j)}\right]+\gamma_j^{-} D\left[\sigma_{-}^{(j)}\right]\right) \rho_{ss}\right]\,,
\end{equation}
providing, \change{ for our example, an objective  function $\langle O\rangle\equiv J_{ss}$} in terms of Pauli strings,
\begin{equation}
\begin{aligned}
J_{ss} &= \frac{\epsilon_h}{2}  (\gamma_h^+ -  \gamma_h^-) - \frac{\epsilon_c}{2}  (\gamma_c^+ -  \gamma_c^-) \\
 &- \frac{\epsilon_h}{2}  (\gamma_h^+ +  \gamma_h^-) \braket{\sigma_h^z}_{ss}  + \frac{\epsilon_c}{2}  (\gamma_c^+ +  \gamma_c^-) \braket{\sigma_c^z}_{ss}\,. \\
\end{aligned}
\end{equation}
This system serves as the first example of how simply ``brute-forcing'' all possible insertions into the \change{adjoint} Lindbladian is often not necessary, as solving with $6$ \textit{automatic} linear constraints results in the exact solution \change{ (i.e., we obtain $\langle O\rangle_{\rm lb}=\langle O\rangle_{\rm ub} $)}, rather than needing to insert all $15$ \change{non-trivial} second-order Pauli strings \cite{consistency, suppmat}.  

Extending this by adding extra qubits between the hot and cold qubits with the same coupling results in a system which appears to be \change{ efficiently solvable}. The automatic linear constraint generation always appears to reach a cycle after $2n^2-n$ constraints, resulting in a full-rank linear system that can be solved exactly. This has been tested up to 100 \changenew{spins} (19900 linear constraints). As such, it \change{was} assumed that this problem is therefore in P since it appears to be solvable in polynomial time. It demonstrates a possible advantage of our constraint set, such that some systems may be shown as being \change{ efficiently} solvable \change{\cite{prosen2008third}}. \change{Note, however, that estimating steady states is known to be hard~\cite{sscomplex} and it is possible to encode the output of any quantum circuit on the steady state of a local master equation~\cite{disscomp}.}

{\it Periodic 1D chain.---}
The next system we consider is that of a periodic $n$-site 1D chain. The Hamiltonian used is that of Ref.\,\cite{hryniuk2024tensor}, the transverse Ising model, specifically a 1D chain such that the final qubit is also coupled to the first qubit and each qubit to a dissipative bath. The qubits are coupled to each other with a \change{coupling strength} $J$, \change{ undergo a transverse field $\eta$} and are connected to the dissipative bath with coefficient $\gamma$. A diagram of this system is given as Fig.\,\ref{fig:merged}(b). The Hamiltonian is as follows, 
\begin{equation}
\label{Hperiodic}
\begin{aligned}
    H &= J \sum_{\braket{i,j}}^n Z_{i}Z_{j} + \change{\eta} \sum_i^n X_i\,, \\
\end{aligned}
\end{equation}
where $\braket{i,j}$ denotes nearest neighbours. To form the full Lindbladian we then use the spin decay operator $\Gamma_k = \frac{\sqrt{\gamma} }{2} (X_k-iY_k)$ such that our full Lindbladian is
\begin{equation}
\label{Lperiodic}
\begin{aligned}
    \lindblad{\rho} &= -i[H, \rho] + \sum_k^n \left( \Gamma_k \rho \Gamma_k^\dagger - \frac{1}{2} \{ \Gamma_k^\dagger \Gamma_k, \rho \} \right) \,. \\
\end{aligned}
\end{equation}
We then consider the case of $n=12$, $J\change{/\gamma}=0.5$, $\change{\eta/\gamma}=0.5$ as given in Ref.\,\cite{hryniuk2024tensor}. In their work they utilize a matrix-product ansatz combined with a Monte-Carlo style optimization. For the 12-site chain, they obtain an estimate for the average magnetization \changenew{$\braket{M}=\frac{1}{n}\sum_i^n \braket{Z_i}$} within $1\%$ of the known optimum in approximately $22$s (based on their Figure 5), whilst we obtain bounds of the optimum $\pm 1\%$ in approximately $50$s\change{: $[-0.821151, -0.801716]$}. Whilst their approximation is faster, our method provides bounds on the quantity in a time of a similar order of magnitude. A table detailing the \changenew{bounded intervals} for the various constraints is given in the End Matter as Table \ref{tab:1Dchain}.

\changenew{To demonstrate the tightness of our bounds irrespective of the observable, we sometimes represent the differences as percentages. These refer to the fraction of the \textit{trivial bounds} that we restrict to. Such bounds are obtained by finding the minimum/maximum of our objective using only the knowledge that each Pauli string should be bounded to $[-1,1]$. For instance, for the magnetization, the trivial bounds are $[-1,1]$.}

\begin{figure}[t]
    \centering
\includegraphics[width=0.9\linewidth]{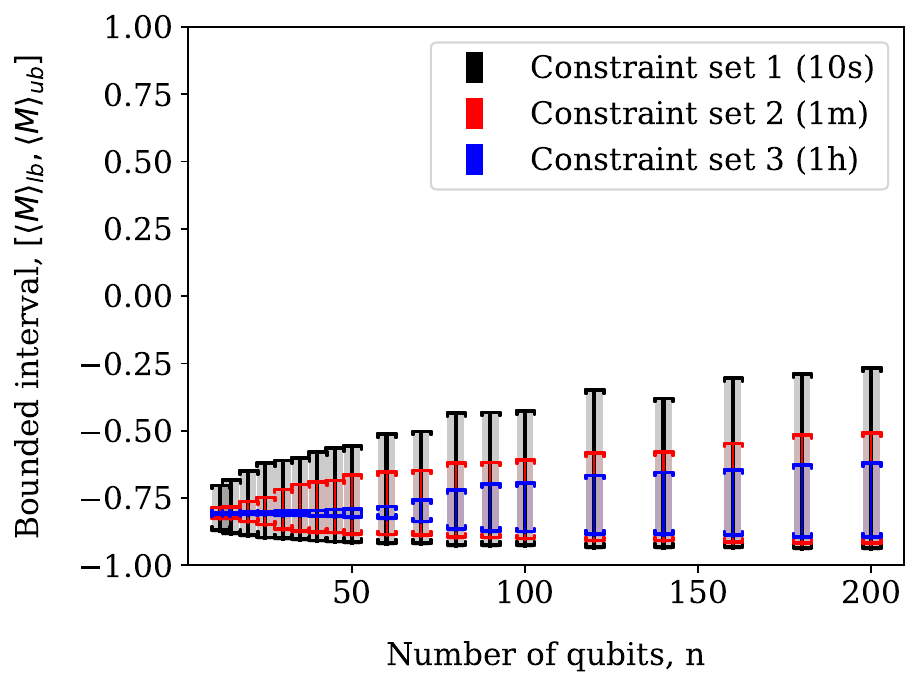}
    \caption{
    \changenew{Upper and lower bounds on the magnetization at the steady state as a function of the system size
    for the periodic linear chain. The plotted intervals represent the certified region in which the true value of the observable lies. The y-axis is scaled to $[-1,1]$ to demonstrate the fraction of the trivial bounds that we restrict to.} We apply method \eqref{sdp-notation} imposing a limited number of linear constraints and moment matrix size (\textit{set 1} - 10000 linear constraints and a $40\times 40$ moment matrix, \textit{set 2} - 30000 constraints and a $100\times 100$ matrix, \textit{set 3} - 70000 constraints and a $250\times 250$ matrix). Symmetry between all qubits is also imposed. The time to optimize per point was roughly the same within each constraint set (\textit{set 1} $\approx 10$ seconds, \textit{set 2} $\approx 1$ minute, \textit{set 3} $\approx 1$ hour).}
    \label{fig:scaling}
\end{figure}

To demonstrate how the method scales with system size for this specific problem, we plot in Fig.\,\ref{fig:scaling} the bounds obtained using \change{a fixed} number of linear constraints and \change{fixed} moment matrix size for a variable number of qubits $n$, ranging from 12 to \change{200}. Although the bounds become worse \change{with increased system size}, they still remain informative in the sense that they significantly restrict the range of possible values. \change{It should be noted that\change{, for instance,} the sign of the observable is still often certified for larger systems.} The computation of all these points required very modest effort and, in particular, the time to optimize is roughly constant within any given constraint set. For instance, for the largest constraint set, in blue, each point takes around \change{1 hour} on a desktop computer \cite{computer} independently of the number of qubits.


{\it {2D ladder}.---}
To explore how our method works in the 2D case, we now consider a 2D ladder system, as shown in Fig.\,\ref{fig:merged}(c); effectively we have two 1D chains linked at each qubit, coupled to a hot and cold bath at either end of the chain. We use \change{an extension of the} Lindbladian and Hamiltonain from Eq.\eqref{eqn:2qubitlimblad}:
\begin{equation}
  \lindblad{\rho}=-i\left[H, \rho\right] +\sum_{j}^n ( \gamma_j^{+} {D}[\sigma_{+}^{(j)}] \rho+\gamma_j^{-} D[\sigma_{-}^{(j)}] \rho ) \,, 
\end{equation}
\begin{equation}
  H = \sum_{i}^n \varepsilon_i \sigma_{+}^{(i)} \sigma_{-}^{(i)}  + g \sum_{\braket{i,j}}^n (\sigma_{+}^{(i)} \sigma_{-}^{(j)}+\sigma_{-}^{(i)} \sigma_{+}^{(j)}) \,,
\end{equation}
\changenew{Here the coefficients $\gamma_j^+$ and $\gamma_j^-$ are only non-zero for the spins connected to the baths.}
Even adding this single extra layer to the system appears to add significant complexity. For the case of bounding the average magnetization of a $2\times5$ grid, with the same computational effort that for the 12-qubit linear chain gave \change{a bound difference} of $1\%$, here gives \change{a bound difference} of $36\%$. \changenew{The largest constraint set tested also reached a bound difference of 36\%  after 2.5 hours \cite{suppmat}. As the bounds do not improve even with larger constraint sets, it seems to suggest that the system has multiple steady states and thus these bounds may be tight.}

{\it Discussion.---}
\label{sec:conc}
We have introduced a novel approach to bound the expectation values of observables in the steady state of open quantum systems. We formulate the problem as an SDP and incorporate various enhancements - linear constraints from the adjoint Lindbladian, moment matrix generation, partial state positivity, and the exploitation of system symmetries. Our method has been tested on relevant models including a two-qubit chain, a one-dimensional chain, and a two-dimensional ladder. 
\changenew{We achieved tight bounds for a 12-site linear chain, as well as \change{informative} bounds for systems composed of hundreds of qubits.}
These findings underscore the effectiveness of our method in providing bounds efficiently, making it \change{a new avenue} for analysing and certifying steady-state properties in open quantum systems.

Looking ahead, there are several directions for future research. 
A promising application is to use the behaviour of \change{our bounds} in terms of a Liouvillian parameter to detect dissipative phase transitions.
\change{Through the application of a fermion-to-qubit mapping one could also apply \change{ our method} to fermionic systems.} 
Finally, investigating the combination of our SDP method with other numerical techniques, such as quantum Monte Carlo, machine learning \cite{requena2023certificates}, or advanced \change{TN} methods, may yield hybrid approaches that capitalize on the strengths of multiple strategies.


\vfill\null

\begin{acknowledgments}
\section{Acknowledgements}

This project has received funding from the European Union’s Horizon 2020 research and innovation programme under the Marie Skłodowska-Curie grant agreement No 847517, 
PNRR MUR Project No. PE0000023-NQSTI,
European Union Next Generation EU PRTR-C17I1,
MICIN and Generalitat de Catalunya with funding from the European Union, NextGenerationEU (PRTR-C17.I1),
the EU projects PASQuanS2.1, 101113690, and Quantera Veriqtas and Compute,
the Government of Spain (Severo Ochoa CEX2019-000910-S and FUNQIP), Fundació Cellex, Fundació Mir-Puig, Generalitat de Catalunya (CERCA program),
the ERC AdG CERQUTE and the AXA Chair in Quantum Information Science.

\vspace{1em}
\noindent
\textbf{Note added:} Upon the finalization of this work, a similar method based on semidefinite programming to bound expectation values of steady-state observables was presented in~\cite{robichon2024bootstrapping} \change{ for} few-mode bosonic systems.

\end{acknowledgments}

\newpage

\vspace{2em}

\newpage

\appendix

\section{  
\large
\textbf{End Matter}
}

\begin{table*}[t]
  \centering
\begin{tabular}{|c|c|c|c|c|c|}
\hline
\textbf{Moment Matrix} & \textbf{Linear} & \textbf{State Reconstruction} & \textbf{Symmetry} & \changenew{\textbf{Bounded Interval}} & \textbf{Time} \\ \hline
None & auto (1000) & None & None & [-1.000000, -0.274680] (36.27\%) & 0.38s \\ \hline
None & auto (10000) & None & None & [-0.926167, -0.670898] (12.76\%) & 22.87s \\ \hline
level 1 (37x37) & auto (10000) & None & None & [-0.882682, -0.697738] (9.25\%) & 9.97s \\ \hline
auto (51x51) & auto (10000) & None & None & [-0.880354, -0.702399] (8.90\%) & 11.70s \\ \hline
None & auto (10000) & all 4-site (794x16x16) & None & [-0.821151, -0.801716] (0.97\%) & 49.49s \\ \hline
auto (51x51) & auto (10000) & all 4-site (794x16x16) & None & [-0.821150, -0.801713] (0.97\%) & 1m13s \\ \hline
None & auto (10000) & all 4-site (794x16x16) & yes (66) & [-0.820893, -0.802149] (0.94\%) & 59.76s \\ \hline
None & auto (30000) & all 4-site (794x16x16) & yes (66) & [-0.817397, -0.805514] (0.59\%) & 2m43s \\ \hline
\end{tabular}
\caption{Bounds on the average magnetisation for the 12-site periodic 1D chain coupled to a bath \change{(Figure \ref{fig:merged} (b)\change{, Eqs.\,\eqref{Hperiodic} and \eqref{Lperiodic})}}, for a variety of possible constraints (the first four columns). \change{The bounds given are the strict lower and upper bounds for the magnetisation in the steady state, whilst the percentages represent the fraction of the full $[-1,1]$ region that the bounds cover.} The other values in parentheses are either the size of matrix or number of constraints added. The system is described by equations (11) and (12), with parameters of $J/\gamma=0.5$, $\eta/\gamma=0.5$.}
\label{tab:1Dchain}
\end{table*}
\section{Appendix A: Density matrix formulation \change{and illustrative examples}}
\label{sec:dm}
\begin{figure}
    \centering
    \includegraphics[width=.33\textwidth]{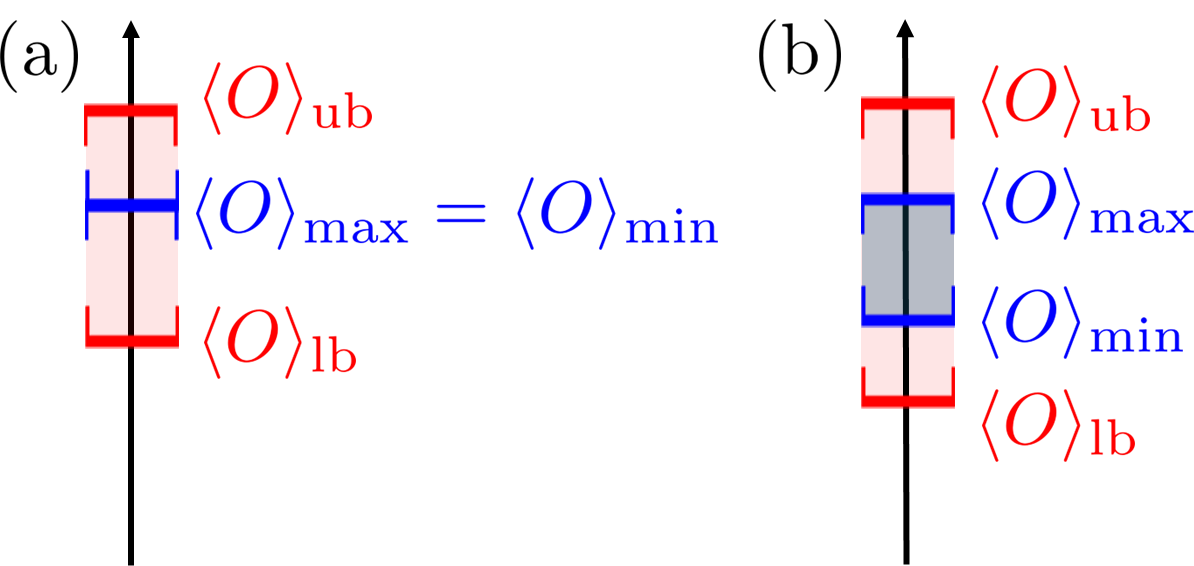}
    \caption{Schematic on the two possible cases: (a) one steady state vs. (b) multiple steady states. $\langle O\rangle_{\max/\min}$ are the outcomes of the exact problem \eqref{sdp-dm-O}.
$\langle {O}\rangle_{\rm ub/lb}$ are the outcomes of problem \eqref{sdp-notation}, which is a relaxation of \eqref{sdp-dm-O}.
In full generality, we have $\langle {O}\rangle_{\rm ub} \geq \langle {O}\rangle_{\max} \geq \langle {O}\rangle_{\min}\geq \langle {O}\rangle_{\rm lb}.$
Observing
$\langle O\rangle_{\max}>\langle O\rangle_{\min}  $ implies the existence of multiple steady states.
}
\label{fig:bounds}
\end{figure}

\change{For small systems (number of qubits $n \lesssim 8$)}, one \change{can} obtain the minimum and maximum value of an observable's expectation value at steady state by solving the following SDP:
\begin{equation}
\label{sdp-dm-O}
\begin{aligned}
\langle O \rangle_{\text{min}/\text{max}}=
& \underset{\,\,\rho\, \text{Hermitian $2^n \times 2^n$}}{\text{min/max}}
& &  {\rm tr}(\rho O) \\
& \text{    \qquad subject to}
&& \mathcal{L}(\rho)=0  \\
&&& \rho \succcurlyeq 0 \\
&&& {\rm tr}(\rho)=1\,.
\end{aligned}
\end{equation}
If the Lindbladian admits only one steady state, then $\langle O \rangle_{\text{min}}=\langle O \rangle_{\text{max}}$, otherwise they \change{ can} differ. Eventually, taking the \textit{argmin} (or \textit{argmax}) an \change{ exact} steady state solution $\rho_{ss}$ can be retrieved, the approach having some connections with recently developed approaches 
on a Hybrid Quantum Processor \cite{PhysRevLett.130.240601}. 
However, solving problem \eqref{sdp-dm-O} 
\change{
suffers from the}
exponential scaling of the density matrix size in the number of qubits $n$.
\change{Problem \eqref{sdp-notation} comes as a scalable relaxation of \eqref{sdp-dm-O} making possible the computation of bounds for bigger systems \change{ (see Fig.\,\ref{fig:bounds})}.
The constraint $\rho \succcurlyeq 0$ can be relaxed via moment-matrix positivity and/or reduced density matrix positivity.
An illustrative example of moment-matrix positivity when setting \change{
$s=5$} and $\change{
(\theta_i)_{i=1}^s}=(\identity, {Y_1}, {Z_1}, {X_2}, {Z_2})\change{
^T}$ in \eqref{eq:def_momentmatrix} is,
%
\begin{equation}
\begin{pmatrix}
    1   & \braket{Y_1} & \braket{Z_1} & \braket{X_2}   & \braket{Z_2} \\
    \braket{Y_1}   & 1 & i\braket{X_1} & \braket{Y_1 X_2}   & \braket{Y_1 Z_2} \\
    \braket{Z_1}   & -i\braket{X_1} & 1 & \braket{Z_1 X_2} &   \braket{Z_1 Z_2} \\
    \braket{X_2}   & \braket{Y_1 X_2 } & \braket{Z_1 X_2 } & 1 &   -i\braket{Y_2} \\
    \braket{Z_2}  & \braket{Y_1 Z_2 } & \braket{Z_1 Z_2 } & i\braket{Y_2}   & 1 \\
\end{pmatrix}
\succcurlyeq 0\,.
\label{eq:moment_matrix-example}
\end{equation}
Note that the presence of the elements $1$ on the diagonal is also effectively a relaxation of the $\tr{\rho}=1$ constraint and that the Pauli reduction rules have been imposed along with commutations of operators acting on different \changenew{spins}.
An example of reduced density matrix positivity on the first two qubits is, instead,
\begin{equation}
\label{reduced}
   \rho_{12} = \frac{1}{4}\sum_{i_1,i_2=0}^3
   \langle \sigma_{i_1}^{(1)}\otimes \sigma_{i_2}^{(2)}\rangle \sigma_{i_1}^{(1)}\otimes \sigma_{i_2}^{(2)}
   \succcurlyeq 0\,,
\end{equation}
with 
$(\sigma_0^{(\alpha)},\sigma_1^{(\alpha)},\sigma_2^{(\alpha)},\sigma_3^{(\alpha)}):=(\identity_\alpha, X_\alpha, Y_\alpha, Z_\alpha)$.
While expression \eqref{reduced} is already in the form $\sum_{\vec\alpha} A_{k,\vec\alpha} \braket{P_{\vec\alpha}} \succcurlyeq 0 $ reported in \eqref{sdp-notation}, we remark that also expression \eqref{eq:moment_matrix-example} can be cast in this form using appropriate matrices $A_{k,\vec\alpha}$. 
}
%


\subsection{Appendix B: Estimating with an ansatz vs. certifying with SDP relaxations}
\label{sec:critique}

We provide a discussion on the difference between estimating with an ansatz and certifying with SDP relaxations. While we focus here on the specific problem of certifying steady-state properties, many considerations of this section can be extended to \change{other settings}.

Let us assume we are interested in the expectation value of an operator $O$
on the steady state(s) of the system, namely on the green convex set {$SS$} in Fig.\,\ref{fig:sets}. 
If the system admits only one steady state the set $SS$ is just a point, hence \change{ implying} a unique \change{
value for $\langle O\rangle_{ss}$}. If there is more than one steady state, one may be interested in estimating the range of possible values, as shown in Fig.\,\ref{fig:bounds}.

{Ansatz-based optimization methods find an approximate estimate \change{$\langle O\rangle_{\rm est}$ }   for the unknown expectation value, by minimizing a suitable cost function within a (generally non-convex) set defined by the ansatz, the blue set $\Omega$ in Figure \ref{fig:sets}(a)-(b).
On this regard, we quote three main facts.
(i) Since the optimization is restricted to a set ${\Omega}$ that is \change{
strictly included in} the set of density matrices (convex set $\cal D$), there can be a deviation from the true value. 
Geometrically, there could be no intersection between $\Omega$ and $SS$, see Fig.\,\ref{fig:sets}(b).
(ii) Even in such smaller set $\Omega$, in general, there is no guarantee to find the global optimum in $\Omega$, namely there is the risk the algorithm can fall and get stuck in \textit{local} minima for the defined cost function (itself sometimes non-convex). This comes as a consequence of the generally non-convex structure of the problem.
{(iii) If the stationary state is not unique, the solution may vary from instance to instance (based on initial conditions, ansatz etc.), even when converging to valid stationary states.}
In conclusion, the estimation provided contains an uncontrolled error, i.e., it is unknown if \change{$\langle O\rangle_{\rm est}$}  overestimates or underestimates the true value and by how much. Its applicability often relies on trusting the assumed ansatz, the ability to escape local minima and so on. 
\changenew{This means that, despite being heuristically informative, such methods do not provide any certainty or certificates.}

On the contrary, convex optimization methods, such as SDPs, 
do not suffer from local minima and they find the \textit{global} optimum of the defined problem \cite{boyd2004convex}, although in many cases this is a relaxation of the original problem.
In our case, {since considering the original convex optimization problem with the whole state as a variable \cite{end-matter-app-dm}
soon becomes intractable (beyond, usually, $\approx 8$ qubits),}
we consider as new variables the {steady-state expectation values, or} moments, and we relax the adjoint Lindbladian constraint and the positivity constraint as described in \change{this work}.
Because it is not guaranteed that a {quantum} state exists satisfying the values of the moments in the new feasible set, by doing such a relaxation we are enlarging the set $SS$ effectively operating on the convex set ${(SS)'}$ that contains the set $SS$, see Fig.\,\ref{fig:sets}(c).
As a consequence, the minimum value found will be smaller than or equal to the true one (see Fig.\,\ref{fig:bounds}). 
This is a lower bound for the unknown quantity $ {\rm tr} (O \rho_{ss})$.
The discussion on obtaining, instead, an upper bound follows an analogous treatment.

From the computational point of view SDPs can be solved efficiently, i.e. in a time that scales polynomially in the number of variables of the problem (the moments in our case) and in the number of constraints provided \cite{vandenberghe1996semidefinite,boyd2004convex}.
%
%
{Estimating with an ansatz \change{(e.g., using TNs, also a scalable tool)} and certifying with SDP relaxations, provide complementary information into a highly complex and relevant problem: 
\change{ considering a system of, say, 100 particles, is it more relevant to get an estimation or certified bounds on a steady-state property of interest? Either answer is valid.}
%
Furthermore, (i) finding with an ansatz a result \change{$\langle O\rangle_{\rm est}$ } not within our SDP bounds \eqref{lbub} certifies that \change{$\langle O\rangle_{\rm est}$ } \change{has an imprecision at least equal to the distance to the closest boundary of our interval},
(ii) finding instead the ansatz-based value \change{$\langle O\rangle_{\rm est}$ } \changenew{within our bounds implies a certified uncertainty for \change{$\langle O\rangle_{\rm est}$ }}.

\change{ Notably,} despite convex optimization methods being widespread powerful certification tools in computer science and engineering \cite{boyd2004convex} and quantum information \cite{navascues2007bounding, navascues2008convergent,skrzypczyk2023semidefinite}, their use in quantum many-body physics/condensed matter is still at its early stage if compared with other approaches, such as estimating under a \change{TN} ansatz.
This opens the way to the introduction of novel methods of great potential. 
Despite this, one can list a series of very recent relevant applications of convex optimization/semi-definite programming in the many-body context, from open-system dynamics \cite{d2024recovering}, ground state property certification \cite{kull2024lower,wang2024certifying}, spectral gap certification \cite{rai2024hierarchy}, Liouvillian gap certification \cite{chen2024boosting}, quantum batteries \cite{salvia2023LE, castellano2025PE},
to generic certification of state convex function values under partial information and \change{ shot noise} \cite{zambrano2024certification}. {Our work is the first to use SDP relaxations to study steady states of open quantum dynamics.}
\begin{figure}[h]
    \centering
    \includegraphics[width=.40\textwidth]{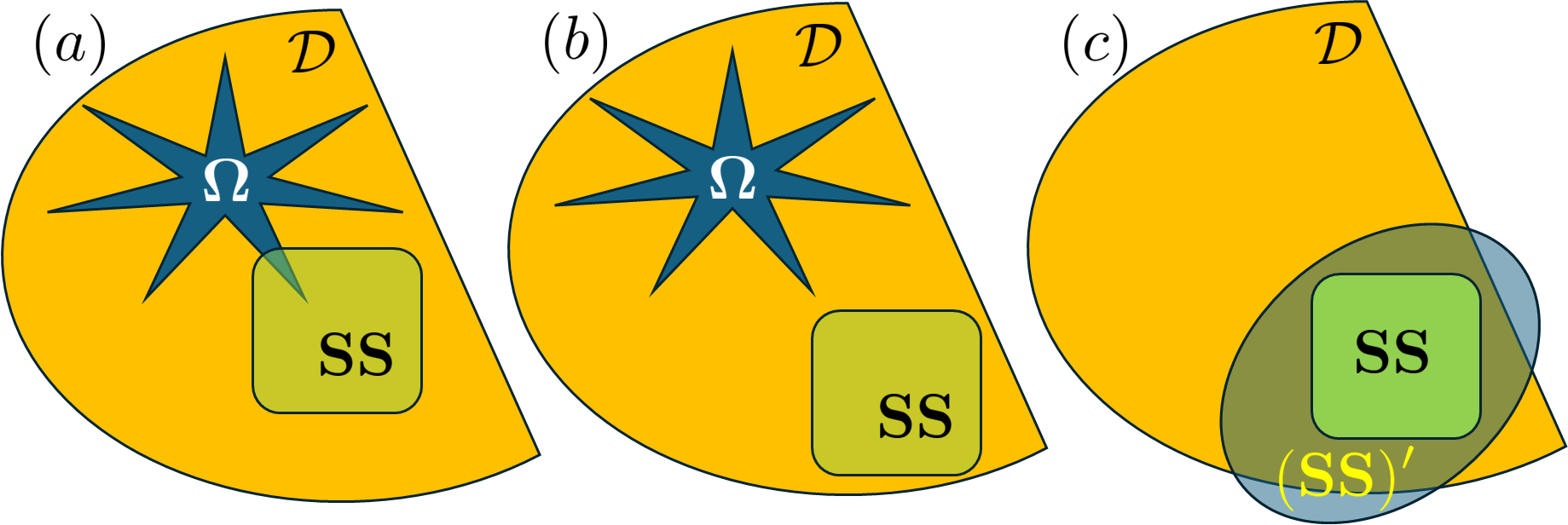}
    \caption{Schematic on the difference between estimating with an ansatz and bounding with an SDP relaxation the steady-state properties of open quantum systems undergoing Lindbladian dynamics. The set $\cal D$ is the convex set of density matrices, that contains the convex set $SS$ of steady states. If the Lindbladian admits a single steady state the set $SS$ is geometrically just a single point. (a)-(b) An estimation method based on an ansatz aims at minimizing a certain cost-function over the set $\Omega$ to find an estimation of ${\rm tr}(O \rho_{ss})$. The latter problem is generally non-convex and the obtained result \change{ contains} an uncontrolled error. (c) An SDP relaxation of the problem implies minimizing (maximizing) on the set $(SS)'\supseteq SS$, finding \textit{global} optima in $(SS)'$. As a consequence the obtained value bounds the possible values ${\rm tr}(O \rho_{ss})$ from below (above).}
    \label{fig:sets}
\end{figure}
}

\newpage
\vspace{.1cm}
\newpage

\bibliography{references, sample}

\begin{thebibliography}{58}
\bibitem [{hba()}]{hbar}%
  \BibitemOpen
  \href@noop {} {\bibinfo  {journal} {In our work, we set $\hbar=1$ and $k_{\rm B}=1$ for simplicity}\ }\BibitemShut {NoStop}%
\end{thebibliography}

\end{document}